\documentclass[12pt]{article}
\usepackage{fullpage}
\usepackage{amsmath,color}
\usepackage{amssymb}
\usepackage{graphicx,psfrag,epsf}
\usepackage{enumerate}
\usepackage{booktabs}
\usepackage{rotating}

\definecolor{marro}{rgb}{0.6,0,0}

\newcommand{\Prob}{\mathbb{P}}

\begin{document}
\bibliographystyle{plain}

\title{\textbf{Bayesian Meta-Analysis of Multiple Continuous Treatments: An Application to Antipsychotic Drugs}}
  \author{Jacob Spertus, Marcela Horvitz-Lennon, and Sharon-Lise Normand}
    \maketitle
    
 \begin{abstract}
 	
 	Modeling dose-response relationships of drugs is essential to understanding their effect on patient outcomes under realistic circumstances. While intention-to-treat analyses of clinical trials provide the effect of assignment to a particular drug and dose, they do not capture observed exposure after factoring in non-adherence and dropout. We develop Bayesian methods to flexibly model dose-response relationships of binary outcomes with continuous treatment, allowing for treatment effect heterogeneity and a non-linear response surface. We use a hierarchical framework for meta-analysis with the explicit goal of combining information from multiple trials while accounting for heterogeneity. In an application, we examine the risk of excessive weight gain for patients with schizophrenia treated with the second generation antipsychotics paliperidone, risperidone, or olanzapine in 14 clinical trials. Averaging over the sample population, we found that olanzapine contributed to a 15.6\% (95\% CrI: 6.7, 27.1) excess risk of weight gain at a 500mg cumulative dose. Paliperidone conferred a 3.2\% (95\% CrI: 1.5, 5.2) and risperidone a 14.9\% (95\% CrI: 0.0, 38.7) excess risk at 500mg olanzapine equivalent cumulative doses. Blacks had an additional 6.8\% (95\% CrI: 1.0, 12.4) risk of weight gain over non-blacks at 1000mg olanzapine equivalent cumulative doses of paliperidone.

 \end{abstract}
 
 \section{Introduction}
 
Decision makers in health care want to know which treatments are most effective and safe. When the treatment in question involves drugs, dosing becomes a key feature to understand. For example, in a randomized controlled trial (RCT) of antipsychotic drugs patients may not fully comply with their assigned treatment, systematically exhibiting poor adherence to a drug that causes adverse outcomes and biasing a naive analysis. If between-trial comparisons are involved, differential prescription patterns and trial length introduce further discrepancies in dosing between drugs. If properly handled, variation in dose is not necessarily problematic: knowledge of the effect of a drug at a given dose may be as valuable to decision makers as the effect of assignment measured in an intention-to-treat (ITT) analysis comparing mean outcomes between assignment groups. Thus, modeling observed dose-response relationships can provide rich, generalizable, and unbiased answers to key questions of interest.

It is as important to understand whether dose-response curves differ among key patient subgroups, a phenomenon known as treatment effect heterogeneity (TEH), in order to enable optimal treatment decisions for diverse patient populations. In schizophrenia research, baseline characteristics like age, gender, and race are likely to moderate the effect of treatment on safety and efficacy \cite{stauffer2010, correll2011}.

In most cases, estimating dose-response relationships entails observational causal inference of one or multiple continuous treatments. Efron and Feldman (1991) discuss compliance-response curves in the presence of a control group. Imai and van Dyk (2004) generalized propensity scores to continuous treatment regimes, while Angrist and Imbens (1995) used instrumental variables to model a continuous measure of treatment intensity \cite{efron1991, imai2004, angrist1995}. Dose-response methods like these that have explicitly adopted a causal inference framework have generally taken a frequentist approach. For reasons discussed further below, Bayesian estimation can be preferable in this setting. At the same time, flexible regression techniques such as splines are preferable as they negate the need to assume linearity of treatment effects. Because dose-response relationships are likely to exhibit ``threshold" or ``plateau" effects, whereby there may be little effect until a certain dose is reached or effects flatten at high exposures, flexible modeling of the curves is essential.

RCT data provide key variables on dosing, outcomes, and potential confounders, but a single RCT is typically not powered for a dose-response analysis, especially in the presence of non-linearity or TEH. Meta-analysis provides a crucial tool to combine information across trials and assess the variation in estimates from different evidence sources\cite{normand1999}. Because it pools information, meta-analysis enables more precise and complex estimates that may be underpowered in a single trial. However, meta-analysis is complicated when modeling flexible dose-response relationships because estimands (typically a smooth function) are simultaneously defined by multiple parameters. In the environmental statistics literature, Dominici et al (2002) used a Bayesian hierarchical spline model to estimate the effect of a single continuous exposure to pollution on mortality across US cities \cite{dominici2002}. Gasparinni et al (2012) built up a frequentist framework for multivariate meta-analysis of temperature and mortality, when the quantity of interest is a spline function defined by multiple parameters \cite{gasparinni2012}. However, these articles did not address inference in the context of multiple treatments and potential TEH, which are key features of drug comparisons. 

To summarize, there are at least 5 major challenges to dose-response estimation with trial data. First, actual cumulative exposure is not randomized and is therefore confounded by baseline variables. Second, non-linear treatment effect curves are defined by multiple parameters simultaneously, complicating inference and the interpretation of results. Third, TEH is crucial to providing guidance for decision makers dealing with diverse populations; because single trials are rarely powered to detect TEH, combining data from multiple trials is imperative. Finally, pooling information from multiple data sources must be accomplished with an approach that can account for between-trial variance.

To address these challenges we combine ideas from causal inference, meta-analysis, and Bayesian inference. We adopt a Bayesian framework using hierarchical models that can naturally incorporate multiple levels of heterogeneity, while posterior parameters of interest can be summarized flexibly and intuitively. This latter point is particularly important in non-linear settings where key elements of the posterior (e.g. dose-response functions) are inherently defined by multiple parameters. A further advantage of the Bayesian approach is intuitive incorporation of substantive information and regularization through the use of priors. To our knowledge, there has been little discussion of Bayesian approaches in the multiple continuous treatment setting, a gap we aim to fill in this paper. 

We adjust for confounding by including controls in an outcome model. We allow for estimation of flexible treatment effect functions through the use of B-splines, where the spline coefficients are distributed hierarchically to partially pool information across trials. Curves can be estimated for multiple distinct treatment drugs. We further define TEH for a binary moderator, allowing separate curves to be estimated for key subgroups. We introduce and discuss some inferential approaches that can provide powerful and intuitive interpretation of posterior results in this complex setting. 

The development of our methods is motivated by an application in psychiatry. Specifically, we seek to understand the effects of second generation antispsychotics (SGAs) on excessive weight gain ($>7\%$ from baseline) under a range of doses. To accomplish this goal we used clinical trial data obtained from the Yale Open Data Access (YODA) Project and Clinical Antipsychotic Trials of Intervention Effectiveness (CATIE), a large trial of SGAs. Actual cumulative exposure was calculated by summing prescribed daily doses over the duration of treatment (accounting for dropout). Adherence data in the form of pill counts or injections was also incorporated when available. Because the potency of SGAs varies, we placed them on the same scale by converting their cumulative doses to doses equivalent to 100mg of olanzapine (OLZ doses), a scaling approach used for comparisons in pyschopharmacology \cite{leucht2014, leucht2015, leucht2016}. Data from placebo arms is incorporated to estimate the intercept (treatment free response) and coefficients of potential confounders with more precision. Further details of our selection and handling of these trials can be found in Spertus et al (2018) \cite{spertus2018}. 

Table 1 summarizes key features of the data. Clearly, the rate of excessive weight gain varies considerably between treatments, but so does actual cumulative exposure. Thus, modeling the dose-response relationship is crucial to understanding the effect of SGAs on weight gain. There are a number of complicating factors, including that some drugs were measured in few trials and multiple baseline variables are likely to confound effect estimates. We are also interested in understanding possible moderating effects of race, which have serious ramifications for clinical practice. Past studies have found conflicting evidence on race driven TEH in terms of efficacy, safety, and prescription patterns \cite{stauffer2010, lieberman2005, ciliberto2005, kuno2002, horvitzlennon2013}. By combining evidence across trials and using flexible hierarchical splines to model the dose-response relationship, we aim to shed further light on this important question. 

\begin{table}[!h]
	\centering
	\caption{Characteristics of data by treatment group. \%ile = percentile; OLZ = olanzapine equivalent dose; PANSS = positive and negative syndrome scale (higher scores imply more severe illness); SD = standard deviation; BMI = body mass index.}
	\begin{tabular}{lrrrr}
		& \multicolumn{4}{c}{Treatment Group} \\ \cline{2-5}
		& \textbf{Placebo} & \textbf{Paliperidone} & \textbf{Olanzapine} & \textbf{Risperidone} \\ 
		\hline
		\# Subjects ($N$) & 1368 & 3482 & 533 & 540 \\
		\# Trials ($J$) & 12 & 13 & 4 & 2 \\
		\hline
		\textit{Outcome and Exposure} & & & & \\
		\hline
		\% $\geq 7\%$ Weight Gain & 4.8 & 10.4 & 17.1 & 11.3 \\
		50th (99th) \%ile 100mg OLZ & 0.0 (0.0) & 6.4 (28.3) & 4.2 (9.0) & 2.6 (23.7) \\
		\hline
		\textit{Baseline Covariates} & & & &\\
		\hline
		Mean Age (SD) & 39.7 (11.9) & 39.4 (11.7) & 38.6 (11.1) & 40.8 (11.8) \\
		Mean PANSS (SD) & 89.0 (15.1) & 89.3 (13.8) & 87.5 (17.1) & 79.8 (14.7) \\
		Mean BMI (SD) & 27.0 (6.4) & 26.9 (6.3) & 27.7 (7.1) & 28.6 (6.2) \\
		\% Female & 37.6 & 38.1 & 31.0 & 34.4 \\
		\% Black & 18.3 & 17.6 & 15.6 & 9.4 \\
		\hline
	\end{tabular}
	\label{tab1:covariates}
\end{table}

 \section{Causal Inference}

 Although treatment assignment is randomized, realized exposure is not, so that we adjust for potential confounding. Our framework for causal inference is based on the Rubin causal model, and borrows ideas from the continuous treatment potential outcomes model formulated in Imai and van Dyk (2004)\cite{imai2004}.  
 
 \subsection{Causal Model and Assumptions}
 
 For subject $i$ in a sample of $n$ subjects, let $\boldsymbol{T}_i$ denote a vector of continuous treatment taking values in a set $\mathcal{T}$, where in our case we have 3 positive valued treatments such that $\mathcal{T} = \mathbb{R}_+^3$. $Y_i \in \{0,1\}$ denotes observed binary outcome and $\boldsymbol{X}_i$ is a vector of baseline covariates. We define the set of potential outcomes $\mathcal{Y} = \{Y_i(\boldsymbol{t}); \boldsymbol{t} \in \mathcal{T}, i = 1...n\}$, where $\boldsymbol{t}$ is an ordinary vector representing a specific realization of the random variable $\boldsymbol{T}$. 
 
 For a given subject we are interested in estimating the function $Y_i(\boldsymbol{t})$ over $\mathcal{T}$. However $\mathcal{T}$ is uncountable when we are dealing with continuous treatment, while we only observe a single value of $Y_i(\boldsymbol{t})$. We therefore assume a degree of homogeneity of treatment effect and smoothness over subjects to generate a meaningfully precise estimate of the potential outcomes. In addition we make the usual assumptions necessary to conduct causal inference.

Stable unit treatment value assignment means that there is only one type of treatment within levels of treatment and that subjects do not interfere with each other, such that a subject's potential outcomes are independent of other subjects' potential outcomes. Positivity means that any subject can (theoretically) receive any level of treatment, i.e: 

\begin{equation}
0 < \Prob(\boldsymbol{T}_i = \boldsymbol{t} | X_i) ~~\forall~~ t, i. 
\end{equation}
With a continuous treatment, positivity is not possible without making smoothness assumptions.

The ignorability assumption is particularly important in observational causal inference. Ignorability implies that conditional on observed covariates potential outcomes are independent of treatment, i.e: 
\begin{equation}
Y_i(\boldsymbol{t}) \perp \boldsymbol{T}_i | \boldsymbol{X}_i ~~\forall~~ \boldsymbol{t}, i
\end{equation}
In a comparison of assigned treatment groups in a randomized trial, randomization helps to ensure ignorability by design. In observational causal inference, it is essential to include all confounders in $\boldsymbol{X}_i$ or estimates will be biased. In contrast to Imai and van Dyk (2004), we do not adjust for confounding by modeling the propensity function $\Prob(\boldsymbol{T}_i = \boldsymbol{t} | X_i)$. We instead adjust for confounding directly in the outcome model \cite{imai2004}.

 \section{A Hierarchical Non-Linear Model}
 \label{sec:model}
 
 \subsection{Notation and Basic Model}
 
Let $j \in \{0,... J\}$ index trials. To assess the causal effect of treatment at varying doses we posit the hierarchical non-linear logistic regression:

\begin{equation}
Y_{ij} \sim \mbox{Bernoulli}\{\mbox{logit}^{-1}[f_j(\boldsymbol{T}_{ij}, \boldsymbol{X}_{ij})]\}.
\end{equation}

This makes for a potentially very high-dimensional problem so we simplify by restraining confounder effects to be additive, linear, and the same across trials. Furthermore, treatment effects are specified to be additive. These two considerations lead to a relatively low dimensional regression: 
$$ f_j(\boldsymbol{T}_{ij}, \boldsymbol{X}_{ij}) = \alpha_{j} + \sum_{k = 1}^{K} f_{jk}(T_{ijk}) + \sum_{p=1}^P \beta_p X_{ijp}.$$

These treatment functions can be flexibly estimated using a B-spline basis. Under this formulation, the exposures for each drug $k$ are expanded into recursively defined power bases with local support \cite{hastie2009}. The bases are defined by boundary points and fixed knots placed within the range treatment. After expansion into a dimension-$L$ B-spline basis with elements $\eta_{kl}(\boldsymbol{T}_{k})$ we express each treatment function as a weighted sum of its basis:

\begin{equation} 
 f_{jk}(T_{ijk}) = \phi_{jk1} T_{ijk} + \sum_{l=2}^L \phi_{jkl} \eta_{kl}(T_{ijk}). \label{eqn:basis_expansion}
\end{equation} 
 
All between trial differences in treatment are expressed through $\phi_{jkl}$. 

To complete the model we specify priors on the confounder coefficients $\beta_p$ and hierarchical priors on the intercept and coefficients of $f_{jk}$. For the confounder coefficients we simply specify the weakly informative prior:
$$ \beta_p \sim t_5(0, 2.5).$$

This implies that, when confounders are binary or scaled to standard normality, most of the prior mass is placed in the interval [-5,5], a sizable range on the log-odds scale\cite{gelman2014}. Let $\boldsymbol{\phi_j} = [\boldsymbol{\phi}_{j1}...\boldsymbol{\phi}_{jK}] = [\phi_{j11},...\phi_{j1L},...\phi_{jK1}... \phi_{jKL}]$, be the length $K \times L$ concatenation of all spline coefficients for each treatment function in trial $j$. Let the overall mean across trials of the intercept be the scalar $\mu_\alpha$, and the mean treatment coefficients be $\boldsymbol{\mu}_\phi = [\mu_{11},...\mu_{1L},...\mu_{K1}... \mu_{KL}]$. The hierarchical prior is specified as:

 \begin{eqnarray}
 \begin{bmatrix}
 \alpha_{j} \\
 \boldsymbol{\phi}_j^{\small\mbox{T}}
 \end{bmatrix} &\overset{{\mbox{iid}}}{\sim}
 \mathcal{N}_{KL + 1}\bigg\{ \begin{bmatrix} 
 \mu_\alpha\\
 \boldsymbol{\mu}_\phi^{\small\mbox{T}}
 \end{bmatrix}, \boldsymbol{\Sigma} \bigg\}  ~~\mbox{where}~  \boldsymbol{\Sigma} = \mbox{D}(\boldsymbol{\sigma}) \boldsymbol{\Omega} \mbox{D}(\boldsymbol{\sigma}) \label{eq:mvn}
 \end{eqnarray}
 
 Here the $(KL + 1) \times (KL + 1)$ covariance matrix $\Sigma$ is decomposed into the product of a diagonal matrix, $\mbox{D}(\boldsymbol{\sigma})$, defined by a $(KL + 1)$ vector of standard deviations $\boldsymbol{\sigma}$, and a $(KL + 1) \times (KL + 1)$ correlation matrix $\boldsymbol{\Omega}$. The entries of $\boldsymbol{\sigma} = [\sigma_{\alpha}, \sigma_{11},...\sigma_{1L},...\sigma_{K1},...\sigma_{KL}]$ represent the between trial variability of each hierarchically specified parameter. Priors for the hierarchical parameters are specified as:
 
 \begin{align}
 \mu_{\alpha} &\sim \mathcal{N}(0,\infty)\\
 \mu_{\phi kl} &\sim t_5(0,2.5)\\ 
 \sigma_{kl} &\sim \mbox{C}^+(0,0.1)\\
 \boldsymbol{\Omega} &\sim \mbox{LKJ}(3)  
 \end{align}
 
 The prior on the intercept  $\mu_{\alpha}$ is completely non-informative. The $t_5(0,2.5)$ priors on $\mu_{\phi kl}$ are made to be weakly informative given the scaling of the exposure (discussed below) \cite{gelman2014}. The half-Cauchy prior on $\sigma_{kl}$ is designed to be slightly regularizing in the sense that it pulls $\sigma_{kl}$ towards zero and thus the trial specific estimates towards their group means, $\mu_\alpha$ or $\mu_{\phi kl}$ \cite{gelman2006}. With $\mbox{C}^+(0,0.1)$ about 95\% of the prior mass is below 1.3, which pools the estimates when there is little information at the trial level but allows for substantial variation if the data are informative.  $\boldsymbol{\Omega}$ is \textit{a priori} distributed according to a Lewandowski, Kurowicka, Joe (LKJ) distribution, determined by a single hyperparameter, which we set equal to 3 to put slightly more prior weight on the identity matrix \cite{Stan2016, burkner2017}. Thus, the covariance matrix $\boldsymbol{\Sigma}$ is slightly regularized through the priors on $\boldsymbol{\sigma}$ and $\boldsymbol{\Omega}$, which is necessary when there is sparsity with respect to $\boldsymbol{\phi}_j$, for example if all drugs are not studied in all trials.

\subsection{Treatment Effect Heterogeneity}

 Treatment effect heterogeneity is accommodated by estimating separate curves within patient subgroups. For example, we could estimate separate dose-response curves for females, black patients, or older patients. Specifically, let the moderators $\bf{M} \subset \bf{X}$ denote a subset of the covariates that we are interested in assessing for possible treatment effect heterogeneity. For simplicity here we posit that $\bf{M} = \boldsymbol{M}$ is a single binary moderator. 
 
The additive treatment effects within levels of $M_{ij}$ are of primary interest. We introduce a new parameter $\boldsymbol{\theta} = [\boldsymbol{\theta}_{1}...\boldsymbol{\theta}_{K}] = [\theta_{11},...\theta_{1L},...\theta_{K1}... \theta_{KL}]$, which parameterizes additional curves for each treatment for subjects with $M_{ij} = 1$, fixed across trials. The models for curves within moderator levels are:

\begin{align} 
f_{jk}(T_{ijk}, M_{ij} = 0) &= \phi_{jk1} T_{ijk} + \sum_{l=2}^L \phi_{jkl} \eta_{kl}(T_{ijk}) \label{eqn:teh_nonblack}\\
f_{jk}(T_{ijk}, M_{ij} = 1) &= \phi_{jk1} T_{ijk} + \theta_{k1} T_{ijk} + \sum_{l=2}^L \big[ \phi_{jkl} \eta_{kl}(T_{ijk}) + \theta_{kl} \eta_{kl}(T_{ijk}) \big]. \label{eqn:teh_black}
\end{align} 

As above, the prior on $\theta_{kl}$ is $t_5(0,2.5)$. Note that $\boldsymbol{\theta}$ does not vary by trial, which implies that the moderating effect is constant across trials. Thus, all trial level heterogeneity is still expressed through $\boldsymbol{\phi}_j$. Also, the basis coordinates $\eta_{kl}(T_{ijk})$ are the same for both subgroups.

For conciseness, we denote the set of all parameters as $\bf {\Upsilon}$. Draws from the posterior are indexed by $q = \{1,...,Q\}$ and ${\bf \Upsilon}_q$ indicates a single draw from the joint distribution of all parameters.

\subsection{Knot Placement and Degrees of Freedom}

This specification assumes fixed knots, which allows for relatively fast computation and simple hierarchical structure. Following Gasparinni et al (2012) \cite{gasparinni2012}, we suggest constructing candidate models of increasing complexity by placing knots symmetrically at fixed quantiles of each treatment: a model with 0 knots is linear in all treatment groups; a 1 knot model is composed of B-splines with a knot at the median; a 2 knot model has knots at .33 and .66, and so on. Boundary knots are placed at dose 0 and the average (across trials) maximum quantile respectively. The lower boundary knot is placed at 0 because it is the natural lower bound for exposure and there will be a high-density of points exactly at zero when there is a placebo arm. We define separate bases for each treatment but they do not vary by trial (as indicated by the `$k$' subscript on the basis terms $\eta_{kl}$ in equation (\ref{eqn:basis_expansion})).    

The final model used for inference can be selected from this candidate set using information criteria. We propose the leave-one-out information criterion (LOO-IC) for fully Bayesian comparisons \cite{vehtari2017}. The LOO-IC can be calculated using Pareto-smoothed importance sampling on a pointwise log-likelihood matrix, a procedure implemented in the \texttt{loo} R package \cite{loo}. The fit of the model sequence is typically concave up in increasing complexity, so a good stopping rule is to fit models with more complexity until the LOO-IC stops decreasing and starts to increase. 
 
 \subsection{MCMC Diagnostics}
 
Posterior draws can be efficiently sampled using Hamiltonian Monte Carlo in Stan and typically only a few thousand draws are needed \cite{stanjstatsoft}. We use the R wrapper package \texttt{brms}, which is a nice interface and has efficient Stan code built in \cite{burkner2017}. The quality of the draws should be assessed by examining trace plots for a few quantities (especially the log-posterior) to spot any pathologies. Convergence can be summarized by the Gelman-Rubin R-hat statistic and all parameters should have an R-hat below 1.1 at convergence \cite{gelman2014}. Graphical and numeric posterior predictive checks provide an additional way to check the model and find places where it does not fit the data well \cite{gelman2014}. 
 
\subsection{Interpretation and Inference}

When the treatment effect is non-linear, the estimand of interest is not a single parameter but a function or multiple functions, each defined by multiple parameters. The best way to assess differences in treatments across doses, treatment types, and potential moderators is to plot the functions defined by the parameters. We can visualize treatment effect heterogeneity by plotting separate functions for each subgroup. We can also evaluate the functions across the joint distribution of parameters and plot lower and upper credible intervals to visualize uncertainty. 

In the logistic regression case, the shape of the function depends somewhat on baseline confounders because terms are not additive after mapping back from the logit to probability scale. Thus the plots give curves for subjects with all baseline confounders set equal to zero. If confounders are appropriately centered and indicators are well chosen, the curves reflect treatment effects for ``typical" subjects. 

Another key quantity of interest is the average treatment effect at fixed doses. We consider the risk difference, $\Delta(k, a)$ which is defined as the expected outcome if all subjects were treated with treatment $k$ at dose $a$ minus the expected outcome if all subjects were treated at no dose. A distribution on the risk difference is obtained by computing it for each draw from the posterior:

$$\Delta_q(k, a) = \mathbb{E}_{\bf X} \{ \mathbb{E}(\boldsymbol{Y} | \boldsymbol{T}_k = a, {\bf X}, {\bf \Upsilon}_q) - \mathbb{E}(\boldsymbol{Y} | \boldsymbol{T}_k = 0, {\bf X}, {\bf \Upsilon}_q) \}.$$

The posterior mean and credible intervals for $\Delta(k, a)$ can be computed from the $Q$ draws obtained.

We summarize comparisons by integrating over a range of doses and averaging over draws. For example, we  calculate the probability that drug 1 has the smallest effect among all drugs over a range of doses $a \in (0,A)$:
\begin{equation}
\Prob\{\Delta(k=1) = \mbox{min}_k \Delta(k)\} =  \frac{1}{Q} \sum_{q = 1}^Q \int_{0}^{A} I \big\{ \Delta_q(k=1, a) = \mbox{min}_k \Delta_q(k, a) \big \} ~\mbox{da} , \label{eqn:ate_comparison}
\end{equation}

The integral can be computed numerically by making a discrete mesh over the range of doses $(0,A)$. This quantity may be useful as a scalar summary of the treatment effects that can be used for decision making. It is straightforward to compute the probability that a drug has the greatest effect merely by taking the maximum instead of the minimum in \ref{eqn:ate_comparison}. 

Treatment effect heterogeneity can also be assessed by taking local average treatment effects across subgroups. Letting $\boldsymbol{M}$ be a binary moderator of interest, we let $\Delta_q(\boldsymbol{M} = 1)$ represent the risk difference in strata defined by $M = 1$ and calculate the distributions within subgroups as:

\begin{align}
\Delta_q(\boldsymbol{M} = 1, a) &=  \mathbb{E}_{\bf X} \{  \mathbb{E}(\boldsymbol{Y} | \boldsymbol{T}_k = a, \boldsymbol{M} = 1, {\bf X}, {\bf \Upsilon}_q) - \mathbb{E}(\boldsymbol{Y} | \boldsymbol{T}_k = 0, \boldsymbol{M} = 1, {\bf X}, {\bf \Upsilon}_q) \} \label{eqn:ate_m1} \\
\Delta_q(\boldsymbol{M} = 0, a) &= \mathbb{E}_{\bf X} \{  \mathbb{E}(\boldsymbol{Y} | \boldsymbol{T}_k = a, \boldsymbol{M} = 0, {\bf X}, {\bf \Upsilon}_q) - \mathbb{E}(\boldsymbol{Y} | \boldsymbol{T}_k = 0, \boldsymbol{M} = 0, {\bf X}, {\bf \Upsilon}_q) \}. \label{eqn:ate_m0} 
\end{align}
 
 Comparisons within drugs but across subgroups are then facilitated by appropriate functions of $\Delta_q(\boldsymbol{M} = 1)$ and $\Delta_q(\boldsymbol{M} = 0)$. A particularly useful function is the difference curve which we define as: $\Delta_q(\boldsymbol{M} = 1, a) - \Delta_q(\boldsymbol{M} = 0, a)$. We simplify computation of the difference curve by ignoring the covariates $\bf X$ which change the intercept. This has the effect of setting all covariates to their default values (assuming centered covariates) and allows computation of a single representative curve instead of separate ones for each subject. The posterior mean and credible intervals are defined at each dose $a$, and the entire difference curve can be plotted. TEH can be assessed by examining the plotted difference curve, where any significant deviation from 0 indicates there may be TEH at that dose. If there is no TEH, the curves are merely offsets of each other and the difference curve will be zero along the plotted range because the intercepts are subtracted off.

\section{Application to OPTICS Trials}

\subsection{Model}

We implemented the model described in section \ref{sec:model}. A linear model provided a baseline for comparison, and more complex models were built using splines and compared. Spline bases were generated for each treatment using the \texttt{bs} R function, with knots placed at appropriate percentiles. These locations were the 50th percentile for a 1 knot basis, 33rd and 66th for a 2 knot basis, 25th, 50th, and 75th for a 3 knot basis, and so on. We built a model for TEH by including race as a binary moderator  variable (black, non-black) as in equations (\ref{eqn:teh_nonblack}) and (\ref{eqn:teh_black}). We tested models of increasing complexity (linear, 1 knot, 2 knots...) using the LOO-IC, and stopped adding knots when the out-of-sample fit worsened.

 \subsection{Diagnostics}
 
 Convergence of our MCMC algorithms was indicated by examining trace plots and the fact that all R-hat statistics were below 1.1. Graphical posterior predictive checks indicated that the model accurately captured the marginal outcome distributions overall, within trials, and within treatment groups. The maximum and minimum outcome proportions across trials were also well captured.

 The LOO-IC (standard error) was 3431 (97) for a linear model, 3416 (97) for a spline model with a single knot, and 3423 (97) for a spline model with 2 knots. Thus LOO-IC indicated that a spline with a single knot provided the best fitting model.

 \subsection{Inference}
 
 Table \ref{tab:outcomes} displays average treatment effects of moving from zero dose (no treatment) to doses equivalent to either 100mg or 500mg of olanzapine for all three treatments studied, along with 95\% credible intervals. It is immediately apparent that higher doses increase the probability of excess weight gain for all three drugs. Olanzapine and risperidone have the largest average effects, but there is considerable uncertainty in their effect sizes as well. 
 
 \begin{table}
 	\centering
 	\begin{tabular}{lrrr}
 		\hline
 		& \textbf{Olanzapine} & \textbf{Paliperidone} & \textbf{Risperidone} \\
 		\hline
 		\textit{Overall} & & & \\
 		\hline
 		0 to 100mg OLZ Equivalents & 1.9 (-0.8, 5.6) & 0.4 (-0.4, 1.1) & 5.8 (-0.5, 13.6) \\
 		0 to 500mg OLZ Equivalents & 15.6 (6.7, 27.1) & 3.2 (1.5, 5.2) & 14.9 (0.0, 38.7) \\
 		\hline
 		\textit{Non-Blacks} & & & \\
 		\hline 
 		0 to 100mg OLZ Equivalents & 1.5 (-0.9, 4.9) & 0.5 (-0.3, 1.2) & 4.9 (-0.7, 12.1) \\
 		0 to 500mg OLZ Equivalents & 15.2 (6.2, 26.0) & 3.1 (1.4, 5.0) & 14.6 (0.3, 37.7) \\
 		\hline
 		\textit{Blacks} & & & \\
 		\hline 
 		0 to 100mg OLZ Equivalents & 4.5 (-3.9, 18.6) & -0.3 (-3.1, 1.8) & 11.5 (-3.0, 34.7) \\
 		0 to 500mg OLZ Equivalents & 17.1 (2.2, 37.3) & 4.1 (-1.2, 9.6) & 16.0 (-5.3, 53.7) \\
 		\hline
 	\end{tabular}
 	\caption{Risk differences for percent probability of 7\% weight gain under various drugs and OLZ equivalent doses. OLZ = olanzapine.}
 	\label{tab:outcomes}
 \end{table}
 
\begin{figure}
 	\includegraphics[width = \textwidth]{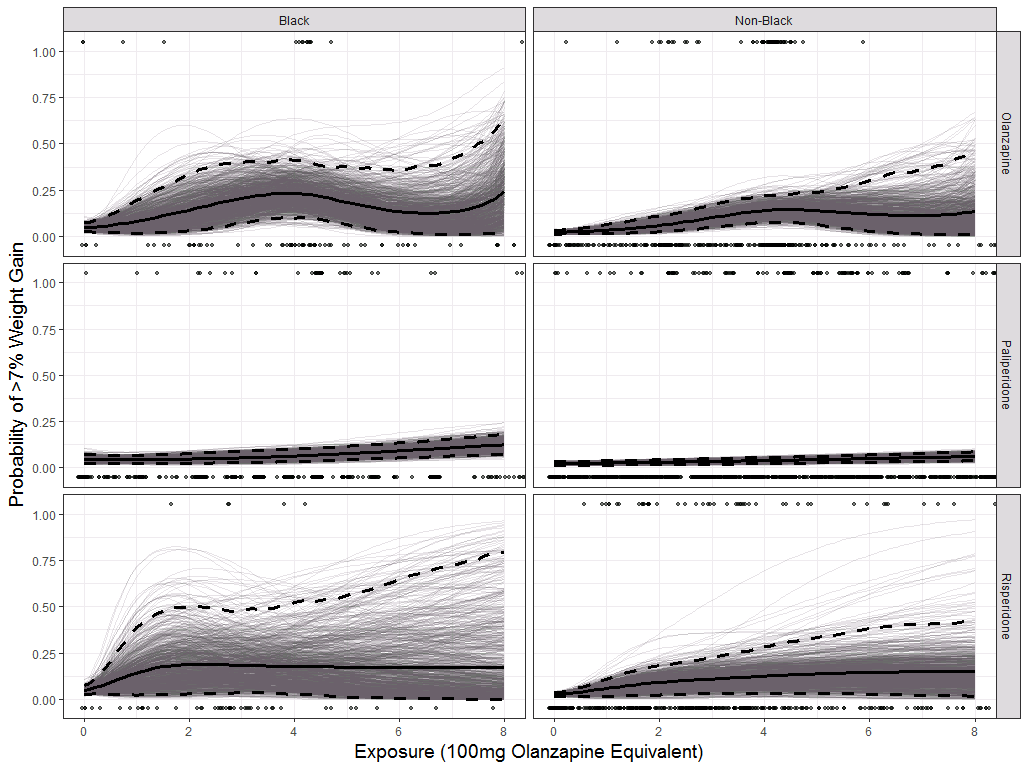}
 	\caption{Draws of dose-response curves for blacks and non-blacks within each drug. Thin grey curves represent a single draw from the posterior, thick black curves are the posterior mean, dotted black lines bound the 95\% credible region for the curves, black points above and below curves mark observed exposures and outcomes. Exposure is on the x-axis and ranges from 0 to 800mg olanzapine equivalent. Y-axis is probability of $>$7\% weight gain. Wide credible intervals for risperidone and olanzapine reflect the fact that they were measured in very few trials and there is evidence inconsistency across trials.}
 	\label{fig:dose_response_curves}
 \end{figure}

 Treatment effects along a range of exposures from 0 to 800mg OLZ equivalents are displayed in figure \ref{fig:dose_response_curves}. Paliperidone has a small positive effect on the risk of weight gain across doses. At high doses, the effect of treatment becomes very uncertain for olanzapine and risperidone, likely due to the small number of subjects at those doses. The effects do seem to level off at high doses of treatment, although the high variance in these dose regions complicates the interpretation of these results. 
 
\begin{figure}
  	\includegraphics[width = \textwidth]{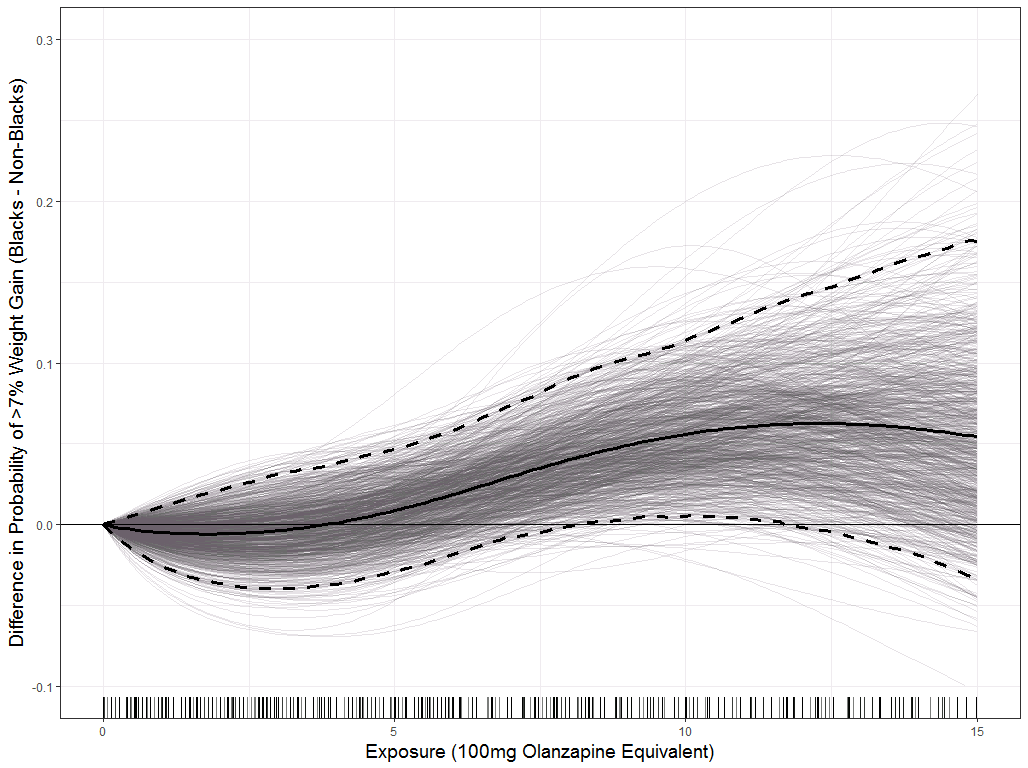}
  	\caption{Draws of TEH for paliperidone showing difference between expected outcome for blacks and non-blacks over a range of exposures to paliperidone (x-axis). Thin grey lines are individual draws from the posterior, solid black line is posterior mean curve, and dashed black lines bound 95\% credible interval. Curves above 0 indicate treatment affects blacks more than non-blacks at that dose, with difference in probability of outcome given on y-axis.}
  	\label{fig:difference_curves}
  \end{figure}

Figure \ref{fig:difference_curves} captures paliperidone TEH over the range of exposure by subtracting the curves for non-blacks from the curves for blacks. If the treatment has a greater effect on blacks, the curve will be positive. We see that the curve deviates significantly from 0 starting at around 700mg OLZ equivalent doses, providing evidence of TEH for blacks at high-doses. Thus, compared with non-blacks, black participants are expected to have an increased weight gain rate of about 5\% on average at 700mg OLZ equivalents or more of paliperidone. For the other two drugs, 95\% credible intervals covered 0 at all doses, indicating no evidence of TEH. Due to the high levels of uncertainty, this is not very informative about whether there is TEH for risperidone or olanzapine.

We used equation \ref{eqn:ate_comparison} to compute the posterior probabilities that each drug was the `best' in that it had the smallest effect on weight gain over a range of exposure from 0 to 500mg OLZ equivalents. All drugs had a high-number of observed cumulative exposures in this range. Probabilities were estimated for both non-black and black subjects but were very similar over this range. Paliperidone had a probability of .85, risperidone had a probability of .06, and olanzapine had a probability of .09 of being the best. Thus the evidence strongly suggests that paliperidone is least likely to cause excessive weight gain at low cumulative doses.
 
\section{Discussion} 
 
 In this paper, we introduced a framework for hierarchical modeling of non-linear treatment effects with potential TEH. Our work was motivated by the need to address non-compliance in RCTs and to gain a better understanding of antipsychotic safety effects across varying cumulative exposure levels. Additionally, we sought to fill a gap in the evidence regarding the moderating effect of race on antipsychotic safety.  Our choice of safety outcome (excessive weight gain) was influenced by the association between excess weight and hypertension, type 2 diabetes, and other risk factors for coronary artery disease. This relationship is especially significant in the context of a rapid growth in SGA utilization in the U.S., partly driven by their frequent use for off-label indications \cite{alexander2011}.
 
 To answer these questions, a meta-analytic framework was needed since trials were not designed to precisely estimate continuous treatment effects or TEH. The model we built incorporated flexible treatment effect modeling through B-splines where the basis coefficients are hierarchically defined to partially pool information across trials. Separate curves can be estimated to assess TEH in the presence of binary moderators. Although we only assessed a single binary moderator, it is straightforward to expand the model to incorporate additional moderators. To assess TEH in a non-linear continuous treatment setting we introduced the concept of difference curves, which allow visualization of the magnitude and uncertainty of effect differences between groups hypothesized to have moderating effects over a range of exposure values. Bayesian computation allow for difference curves to be computed quickly and simply using posterior draws. Although our approach was motivated by the need to assess SGA effects, it can generalize readily to other drug trials and applications outside psychiatry where one or more continuous treatments are of interest. 
 
 Posterior predictive checks indicated that our model captured outcome distributions well across trials and treatments. Comparison using LOO-IC favored a spline model with a single knot, accounting for race as a binary moderator.
 
 All drugs led to increased probability of weight gain over their range of exposures, though olanzapine and risperidone provided very imprecise estimates of effect curves. There was some evidence of a leveling off of the effects at higher doses, a feature that we were able to capture using a non-linear model. In addition, we found evidence that race moderates the paliperidone's association with weight gain, with increased risk of weight gain for black subjects. Nevertheless, paliperidone was likely the best among the alternatives we considered for both blacks and non-blacks. We were unable to find any realistic level of moderation of race for risperidone or olanzapine due to high levels of uncertainty.

 A major challenge in our application, and one that is likely to arise in other multivariate continuous treatment designs, is how to standardize exposures. We chose to use olanzapine equivalent doses, but other absolute scales like chlorpromazine equivalents or recommended daily doses are also possible. Another possibility is to scale the treatments relatively, e.g. divide the treatment variables so 100 is always the maximum dose, but this could introduce problems if certain drugs were included in the original trials at relatively higher doses. In \cite{spertus2018}, we found results to be relatively robust to the specific scaling used. Another challenge was that relatively few trials assessed olanzapine and risperidone (4 and 2 respectively), leading to strong uncertainty in these estimates. A potential solution is to fully pool the data for these drugs and not model them hierarchical, but this may bias inference by ignoring trial-level heterogeneity. As more individual participant level trial data is made available, it will be possible to use our framework with additional information to make more precise inferences on more drugs and better understand optimal treatments for diverse subjects.

 \section{Acknowledgments} 
  
 This study was approved by the IRB of the Harvard Faculty of Medicine. We would like to thank our collaborators at the Harvard Catalyst Reactor Program, Hardeep Ranu and Gary Gray; Marsh Wilcox, Janssen Scientific Director \& Fellow; Linda Valeri, Harvard Medical School; and John Jackson, Johns Hopkins Bloomberg School of Public Health. 
 
 \section{Funding}
 
 Mr. Spertus' effort was conducted with support from The Harvard Clinical and Translational Science Center (National Center for Research Resources and the National Center for Advancing Translational Sciences, National Institutes of Health Award UL1 TR001102),  Harvard Catalyst, and Harvard University. Drs. Horvitz-Lennon and Normand were funded by Harvard Catalyst and the National Institute of Mental Health (R01-MH106682). Dr. Normand was also supported by the National Institute of General Medical Sciences (R01-GM111339). The content is solely the responsibility of the authors and does not necessarily represent the official views of Harvard Catalyst, Harvard University and its affiliated academic healthcare centers, or the National Institutes of Health. 
 
 \section{Data Availability}
 This study, carried out under YODA Project No. 2015-0678, used data obtained from the Yale University Open
 Data Access Project, which has an agreement with Janssen Research \& Development, L.L.C.. The interpretation and reporting of research using this data are solely the responsibility of the authors and does not necessarily represent the official views of the Yale University Open Data Access Project or Janssen Research \& Development, L.L.C. The CATIE data used in this paper reside in the NIH-supported NIMH Data Repositories [NIMH Data Repositories DOI: 10.15154/1373363]; Principal Investigators of original data: J. Lieberman (N01-MH090001) and P. Sullivan (R01-MH074027). 
 
 The CATIE data is available from the National Institute of Medical Health Repository and Genomics Resource (https://bioq.nimhgenetics.org/studies/?studyId=20). The Janssen Trials are available from the Yale Open Data Access Project (http://yoda.yale.edu/multiple-ncts-optics-trial-bundle).  
  
\bibliography{OPTICS_stats_bibliography}

\begin{thebibliography}{10}

\bibitem{alexander2011}
G.~Caleb Alexander, Sarah~A Gallagher, Anthony Mascola, Rachael~M Moloney, and
  Randall~S Stafford.
\newblock Increasing off-label use of antipyschotic medications in the united
  states, 1995-2008.
\newblock {\em Pharmacoepdemiol Drug Saf}, 20:177--184, 2011.

\bibitem{angrist1995}
Joshua~D Angrist and Guido~W Imbens.
\newblock Two-stage least squares estimation of average causal effects in
  models with variable treatment intensity.
\newblock {\em Journal of the American Statistical Association}, 90:431--442,
  1995.

\bibitem{burkner2017}
Paul-Christian Burkner.
\newblock brms: an r package for bayesian multilevel models using stan.
\newblock {\em Journal of Statistical Software}, 80, 2017.

\bibitem{stanjstatsoft}
Bob Carpenter, Andrew Gelman, Matt Hoffman, Daniel Lee, Ben Goodrich, Michael
  Betancourt, Michael~A Brubaker, Jiqiang Guo, Peter Li, and Allen Riddell.
\newblock Stan: a probabilistic programming language.
\newblock {\em Journal of Statistical Software}, 76, 2017.

\bibitem{ciliberto2005}
Natalie Ciliberto, Cynthia~A Bossie, Ronald Urioste, and Robert~A Lasser.
\newblock Lack of impact of race on the efficacy and safety of long-acting
  risperidone versus placebo in patients with schizophrenia or schizoaffective
  disorder.
\newblock {\em International Clinical Psychopharmacology}, 20:207--212, 2005.

\bibitem{correll2011}
Christoph~U Correll, Todd Lencz, and Anil~K Malhotra.
\newblock Antipsychotic drugs and obesity.
\newblock {\em Trends Mol Med}, 17:97--107, 2011.

\bibitem{dominici2002}
Francesca Dominici, Michael Daniels, Scott~L Zeger, and Jonathan~M Samet.
\newblock Air pollution and mortality: estimating regional and national
  dose-response relationships.
\newblock {\em Journal of the American Statistical Association}, 97:100--111,
  2002.

\bibitem{efron1991}
B~Efron and D~Feldman.
\newblock Compliance as an explanatory variable in clinical trials.
\newblock {\em Journal of the American Statistical Association}, 86:9--17,
  1991.

\bibitem{gasparinni2012}
Antonio Gasparinni, B~Armstrong, and M~G Kenward.
\newblock Multivariate meta-analysis for non-linear and other multi-parameter
  associations.
\newblock {\em Statistics in Medicine}, 31:3821--3839, 2012.

\bibitem{gelman2006}
A~Gelman.
\newblock Prior distributions for variance parameters in hierarchical models.
\newblock {\em Bayesian Analysis}, 1:515--533, 2006.

\bibitem{gelman2014}
A~Gelman, JB~Carlin, HS~Stern, DB~Dunson, A~Vehtari, and DB~Rubin.
\newblock {\em Bayesian Data Analysis, Third Edition}.
\newblock CRC Press, Boca Raton, FL, 2014.

\bibitem{hastie2009}
Trevor Hastie, Robert Tibshirani, and Jerome Friedman.
\newblock {\em The Elements of Statistical Learning: Second Edition}.
\newblock Springer, 2009.

\bibitem{horvitzlennon2013}
Marcela Horvitz-Lennon, Julie~M Donohue, Judith~R Lave, Margarita Alegria, and
  Sharon-Lise~T Normand.
\newblock The effect of race-ethnicity on clozapine outcomes among medicaid
  beneficiaries with schizophrenia.
\newblock {\em Psychiatric Services}, 64:230--237, 2013.

\bibitem{imai2004}
Kosuke Imai and David~A van Dyk.
\newblock Causal inference with general treatment regimes: Generalizing the
  propensity score.
\newblock {\em Journal of the American Statistical Association}, 99:854--866,
  2004.

\bibitem{kuno2002}
Eri Kuno and Aileen~B Rothbard.
\newblock Racial disparities in antipyschotic prescription patterns for
  patients with schizophrenia.
\newblock {\em The American Journal of Psychiatry}, 159:567--572, 2002.

\bibitem{leucht2016}
Stefan Leucht, Myrto Samara, Stephan Heres, and John~M Davis.
\newblock Dose equivalents for antipyschotic drugs: the ddd method.
\newblock {\em Schizophrenia Bulletin}, 42:90--94, 2016.

\bibitem{leucht2015}
Stefan Leucht, Myrto Samara, Stephan Heres, Maxine~X Patel, Toshi Furukawa,
  Andrea Cipriani, John Geddes, and John~M Davis.
\newblock Dose equivalents for second-generation antipsychotic drugs: the
  classical mean dose method.
\newblock {\em Schizophrenia Bulletin}, 41:1397--1402, 2015.

\bibitem{leucht2014}
Stefan Leucht, Myrto Samara, Stephan Heres, Maxine~X Patel, Scott~W Woods, and
  John~M Davis.
\newblock Dose equivalents for second-generation antipsychotic drugs: the
  minimum effective dose method.
\newblock {\em Schizophrenia Bulletin}, 40:314--326, 2014.

\bibitem{lieberman2005}
JA~Lieberman, TS~Stroup, JP~McEvoy, MS~Swartz, RA~Rosenheck, DO~Perkins,
  RS~Keefe, SM~Davis, CE~Davis, BD~Lebowitz, J~Severe, JK~Hsiao, and Clinical
  Antipsychotic~Trials of~Intervention Effectiveness (CATIE)~Investigators.
\newblock Effectiveness of antipsychotic drugs in patients with chronic
  schizophrenia.
\newblock {\em The New England Journal of Medicine}, 353:1209--1223, 2005.

\bibitem{normand1999}
Sharon-Lise Normand.
\newblock Meta-analysis: Formulating, evaluating, combining, and reporting.
\newblock {\em Statistics in Medicine}, 18:321--359, 1999.

\bibitem{spertus2018}
Jacob Spertus, Marcela Horvitz-Lennon, Haley Abing, and Sharon-Lise Normand.
\newblock Risk of weight gain for specific antipyschotic drugs: a bayesian
  network meta-analysis of individual participant level clinical trial data.
\newblock {\em npj Schizophrenia}, In Press, 2018.

\bibitem{stauffer2010}
Virginia~L Stauffer, Jennifer~L Sniadecki, Kevin~W Piezer, Jennifer Gatz, Sara
  Kollack-Walker, Vicki~Poole Hoffmann, Robert Conley, and Todd Durell.
\newblock Impact of race on efficacy and safety during treatment with
  olanzapine in schizophrenia, schizophreniform, or schizoaffective disorder.
\newblock {\em BMC Pyschiatry}, 10, 2010.

\bibitem{Stan2016}
Stan~Development Team.
\newblock Stan modeling language users guide and reference manual, 2016.
\newblock Version 2.14.0.

\bibitem{vehtari2017}
Aki Vehtari, Andrew Gelman, and Jonah Gabry.
\newblock Practical bayesian model evaluation using leave-one-out
  cross-validation and waic.
\newblock {\em Statistics and Computing}, 27:1413--1432, 2017.

\bibitem{loo}
Aki Vehtari, Andrew Gelman, Jonah Gabry, Juho Piironen, and Ben Goodrich.
\newblock Package 'loo'.
\newblock 2017.

\end{thebibliography}

\appendix

\section{Posterior Predictive Checks}

To check the fit of our model to the data, we tested the proportion of subjects experiencing the outcome within treatment groups and the maximum/minimum proportion of outcomes across trials. Both checks indicated the model fit the data well.

\begin{figure}
  	\includegraphics[width = \textwidth]{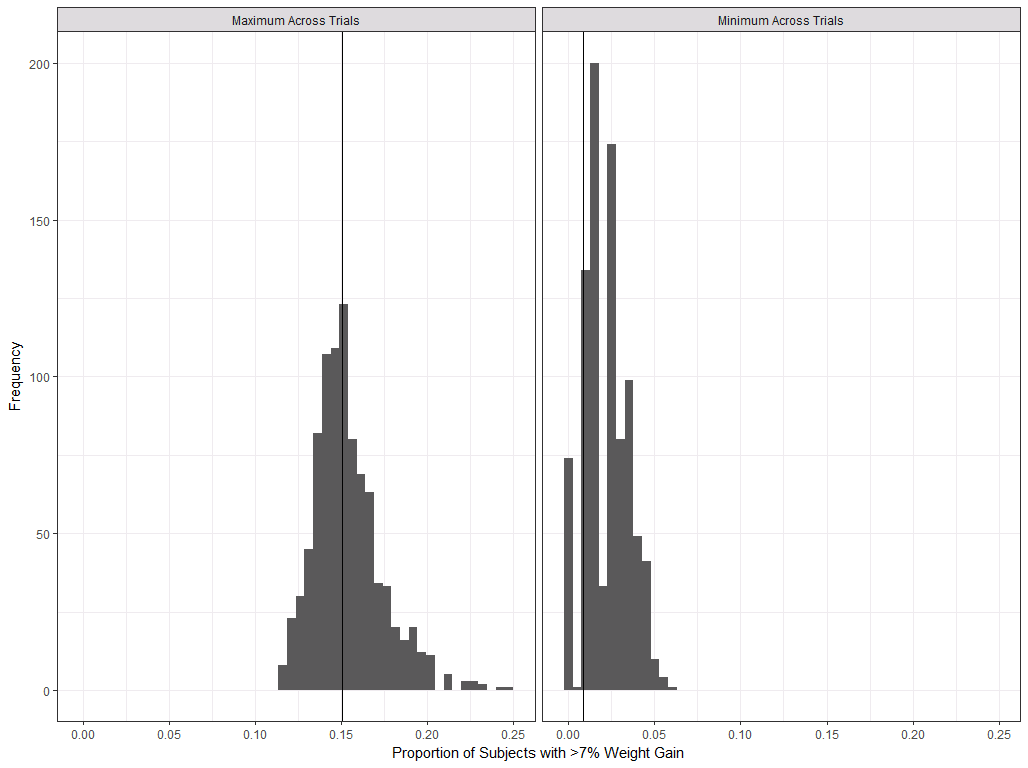}
  	\caption{Posterior predictive distributions. Histograms show posterior predictive draw of the maximum and minimum outcome proportion across trials, vertical lines are observed values.}
  	\label{fig:trial_ppcheck}
\end{figure}

\begin{figure}
	\includegraphics[width = \textwidth]{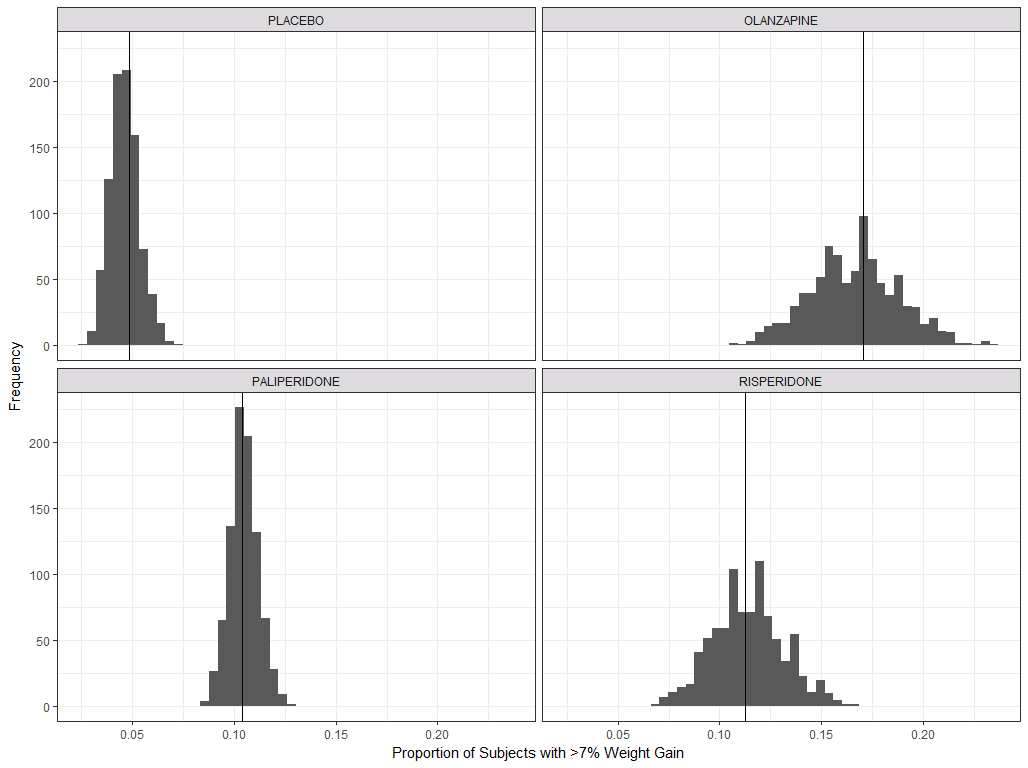}
	\caption{Posterior predictive distributions showing proportions of outcomes in each treatment group. Histograms show posterior predictive draws, vertical lines are observed values.}
	\label{fig:treatment_ppcheck}
\end{figure}

\end{document}